\documentclass{article}

\usepackage[margin=1in]{geometry}
\usepackage{amssymb}
\usepackage{siunitx}
\usepackage{xcolor}
\usepackage{graphicx}

\title{A new sensing paradigm for the vibroacoustic detection of
pedicle screw loosening}
\author{Matthias Seibold$^{1,2}$, Bastian Sigrist$^1$, Tobias Götschi$^3$,\\ 
Jonas Widmer$^3$, Sandro Hodel$^4$, Mazda Farshad$^4$,\\ 
Nassir Navab$^2$, Philipp Fürnstahl$^1$, Christoph J. Laux$^4$}
\date{%
    $^1$Research in Orthopedic Computer Science, Balgrist University Hospital, University of Zurich, Switzerland \\
    $^2$Computer Aided Medical Procedures, Technical University Munich, Germany\\%
    $^3$ Spine Biomechanics, Balgrist University Hospital, ETH Zurich, Switzerland \\
    $^4$ Department of Orthopedics, Balgrist University Hospital, University of Zurich, Switzerland
    \\[2ex]%
}

\begin{document}

\maketitle

\begin{abstract}

There is an unmet clinical need for developing novel methods to complement and replace the current radiation-emitting imaging-based methods for the detection of loose pedicle screws as a complication after spinal fusion surgery which fail to identify a substantial amount of loose implants. In this work, we propose a new methodology and paradigm for the radiation-free, non-destructive, and easy-to-integrate detection of pedicle screw loosening based on vibroacoustic sensing. Furthermore, we propose a novel simulation technique for pedicle screw loosening, which is biomechanically validated. For the detection of a loose implant, we excite the vertebra of interest with a sine sweep vibration at the spinous process and attach a custom highly-sensitive piezo contact microphone to the screw head to capture the propagated vibration characteristics which are analyzed using a detection pipeline based on spectrogram features and a SE-ResNet-18. To validate the proposed approach, we conducted experiments using four human cadaveric lumbar spine specimens and evaluate our algorithm in a cross validation experiment. Our method reaches a sensitivity of $91.50 \pm 6.58 \%$ and a specificity of $91.10 \pm 2.27 \%$. The proposed system shows great potentials for the development of alternative assessment methods for implant loosening based on vibroacoustic sensing. 

\end{abstract}

\section{Introduction}

Spinal instrumentation with pedicle screws is an broadly established and increasingly used intervention in the surgical treatment of degenerative diseases, injuries, deformities or tumors of the spine. \cite{mobbs2015fusion,rajaee2012spinal,kobayashi2022trends}. Hereby, the spinal segment is stabilized by driving screws into both pedicles of the respective vertebra and connect them with rods on either side that absorb most of the biomechanical forces. One of the most common postoperative complications of this surgical procedure is screw loosening, which  is often associated with persistent pain and therefore eventually requires revision surgery. Pedicle screw loosening usually manifests itself in a fan-shaped cavity around the screw shaft and results in screw toggling, therefore allowing movement between the instrumented segments \cite{liebsch2018pedicle,law1993mechanisms,baluch2014screw}
The risk of pedicle screw loosening has been reported in literature as $1-3 \%$ per screw and 12.3\% per patient \cite{bredow2016pedicle}. In osteoporotic bone there is an even higher risk of pedicle screw loosening in the range of $50-60 \%$ \cite{elsaman2013reduced,kim2020clinical} which imposes a highly relevant clinical problem in an ageing population \cite{hoppe2017pedicle}.

When patients report implant-related pain or instabilities, the standard way to asses potential implant loosening is to employ different medical imaging modalities, such as Magnetic Resonance Imaging (MRI), Computed Tomography (CT), or planar radiographs. In a prospective clinical study, Spirig et al. found the sensitivity and specificity in detection of screw loosening to be 43.9\% and 92.1\% for MRI, 64.8\% and 96.7\% for CT, and 54.2\% and 83.5\% for standard radiographs, respectively \cite{spirig2019value}. In clinical practice, CT remains the gold standard for the assessment of pedicle screw loosening but fails to detect a substantial amount of loose implants and exposes the patient to radiation, even though it is recommended to use low-dose CT protocols if possible \cite{abul2014ct}. Therefore, there is a clinical need to develop alternative non-invasive and radiation-free methods for the detection of loose pedicle screws and better understand their clinical correlation. 

After screw loosing was diagnosed by imaging, a loose screw is confirmed intraoperatively by measuring a low torque when removing the screw \cite{spirig2019value,wu2019pedicle} which destroys the bone-implant interface in tight screws and further weakens the spine segment, if the screw was incorrectly identified as loose. Therefore, a reliable and non-invasive method for the intra-operative assessment of pedicle screw loosening for the intraoperative use in revision surgery would be highly desirable. 

Acoustic sensing is a non-invasive, radiation-free and easy-to-integrate modality which has been shown to have great potential for various medical applications such as intraoperative tissue classification \cite{ostler2020acoustic,illanes2018novel}, surgical error prevention \cite{seibold2021realtime}, or patient monitoring \cite{romero2019sleep}. Acoustic emission analysis has furthermore been employed in the condition assessment and early diagnosis of orthopedic implants, but has mainly been applied to artificial hip and knee joints so far \cite{khokhlova2021review}. Schwarzkopf et al. recorded the acoustic emissions of different types of knee implants using a handheld measurement system. The analysis of the data revealed correlations to the implant status and time from implantation \cite{schwarzkopf2011ae}. Rodgers et al. proposed a system for monitoring the acoustic emissions of THA implants and characterized the squeaking of hard-on-hard bearing surface combinations \cite{rodgers2014emission}. Fitzpatrick et al. developed a monitoring system based on acoustic emission sensing to measure the wear of total hip replacement implants and compared the frequency characteristics of in-vivo and in-vitro recordings \cite{fitzpatrick2017wear}.
For the assessment of implant stability, a proof-of-concept-study was published by Ewald et al., who developed a prototype and simulator-based experimental setup for the detection of total hip replacement implant loosening using an acoustic sensor system \cite{ewald2011acoustic}. Arami et al. developed a vibroacoustic system for ex vivo detection of loosening of total knee replacement implants. They applied harmonic vibration to the tibia and measured the resulting vibrations on the implant surface using an accelerometer \cite{arami2018knee}. 

The systems described above employ frequency analysis or classical signal processing methods to define thresholds or describe the characteristics of frequency components. However, as deep learning-based methods have recently replaced and outperformed classical approaches for solving audio specific tasks such as speech recognition \cite{nassif2019speech} and environmental sound processing \cite{miyazaki2019environmental}, these techniques have also successfully been applied to medical applications \cite{romero2019sleep,seibold2021realtime,xu2021listen,seibold2021femoralstem}.

In this work, we propose a novel method to assess the hold of pedicle screws based on vibroacoustic sensing. In the first step, we developed an experimental approach to simulate pedicle screw loosening in human cadaveric specimens. We instrumented four human cadaveric lumbar spine specimens and validated the screw loosening by analyzing the relative movement between implant and instrumented vertebra in fixed and loose configurations using a biomedical testing machine and an optical tracking system. For the detection of screw loosening, we excite the anatomy by using a vibration device to send a sine sweep into the bone and measure the propagated vibrations directly at the screw head. Subsequently, we developed an automated algorithm based on log-mel spectrograms and a SE-ResNet-18 to detect screw loosening based on the characteristics of the captured signal and thoroughly evaluate the performance of the proposed algorithm in a leave-one-specimen-out cross validation experiment. The proposed proof-of-concept system can be directly used for the intraoperative assessment of pedicle screw loosening during revision surgery. 


\section{Material and methods}

\subsection{Experimental approach for the simulation of Pedicle Screw Loosening}
\label{sec:loosening}

Pedicle screw loosening is usually simulated in biomechanical experiments by applying dynamic loading over thousands of cycles in experimental setups \cite{liebsch2018pedicle,choma2011pedicle,kueny2014screw}. As we are not interested in measuring the biomechanical forces but rather simulate a loose implant in terms of relative movement between target anatomy and implant (screw toggling), we developed a different approach capable of simulating the mechanics of pedicle screw loosening in a faster way. To this end, CT scans of the cadaver specimens were acquired and the screw entry points and target angles were planned by an experienced spine surgeon according to the standard clinical routine in a 3D surgical planning software (CASPA, University Hospital Balgrist, Switzerland).

\begin{figure}
    \centering
    \includegraphics[width=0.7\textwidth, angle=180]{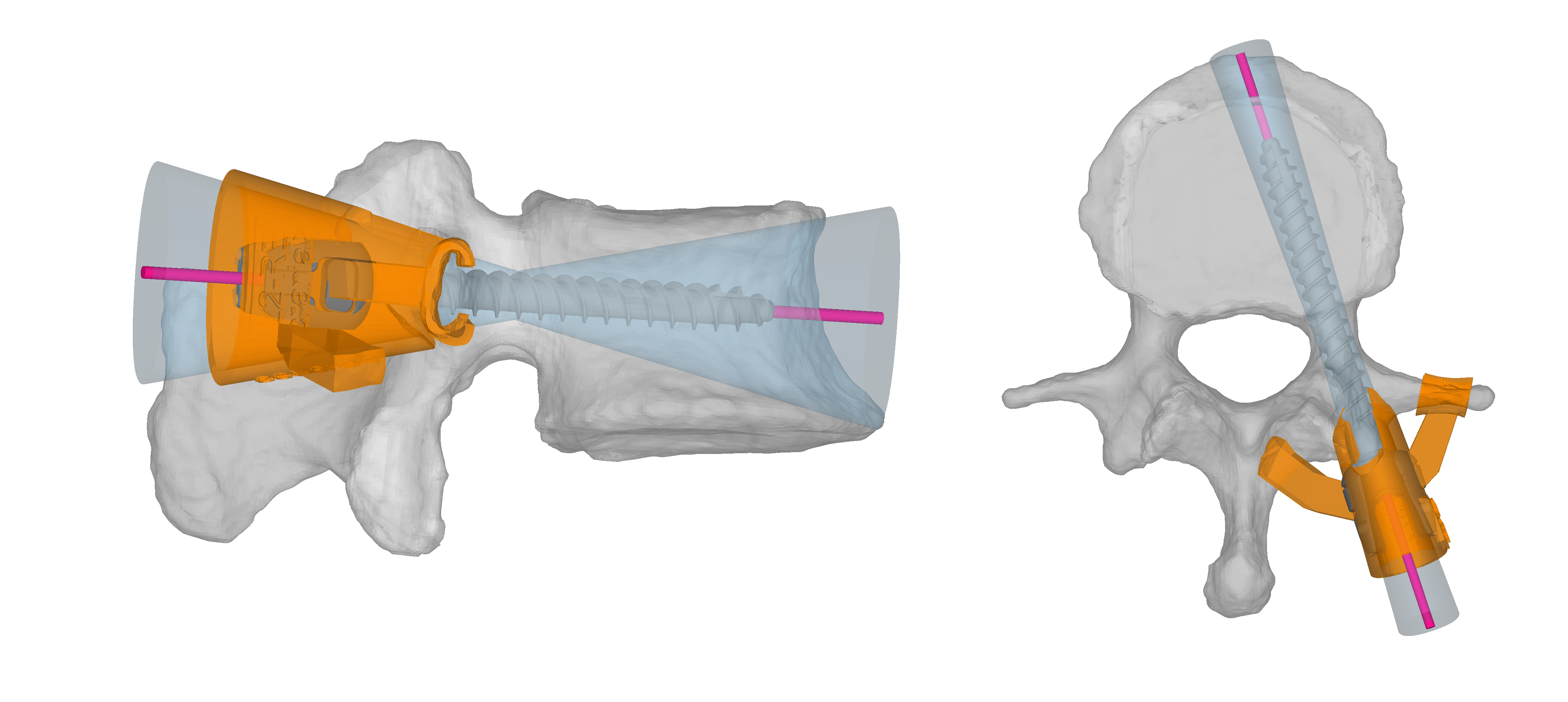}
    \caption{Instrumentation of the cadaveric spine specimens and simulation of the pedicle screw loosening using the 3D printed drill guide. On the left side, the sagittal view with an angle deviation of \SI{25}{\degree} is illustrated; the right side shows the transversal view with with an angle deviation of \SI{5}{\degree}. The custom 3D printed drill guide is colored in orange, the planned screw trajectory in pink, and the fan-shaped cavity in light blue.}
    \label{fig:drilling}
\end{figure}

A two-step data acquisition procedure was performed were first stable screws were inserted and measured (control group) and, in a second step, screw loosening was simulated (intervention group). In the first step, we inserted Medacta pedicle screws (Medacta, Castel San Pietro, Switzerland) along the planned screw trajectory using a classical approach without predrilling. In the second step, the screws were intentionally loosened. Therefore, we used the planned screw trajectories to design custom 3D printed drill guides, as illustrated in figure \ref{fig:drilling}, to drill a fan-shaped hole with an angle deviation of \SI{25}{\degree} in the sagittal and \SI{5}{\degree} in the transversal plane into the respective vertebra. The CTs were manually segmented and the drill guides leveraged the concept of patient specific instruments (PSI) \cite{hafez2017psi} where the undersurface of the guide is shaped as a negative of the target anatomy surface, therefore only fitting in one unique position on the vertebrae. All custom drill guides were manufactured using a highly accurate laser sinter 3D printer (EOS Formiga P396, EOS Systems, Krailling, Germany).

For all four lumbar spine specimens, the vertebrae L2 and L4 were instrumented and only the screws in vertebra L2 were intentionally loosened in a second step. This approach allows to maintain a fixation on one implant side (L4) to measure the relative movement between implant and target vertebra (L2) during movement using an optical tracking system. All surgical steps were performed by an experienced spine surgeon and a bilateral posterior approach through the Wiltse interval was chosen to preserve the skin directly over the spinous processes. 

\subsection{Validation of Pedicle Screw Loosening}
\label{sec:validation}

\begin{figure}
    \centering
    \includegraphics[width=0.5\textwidth]{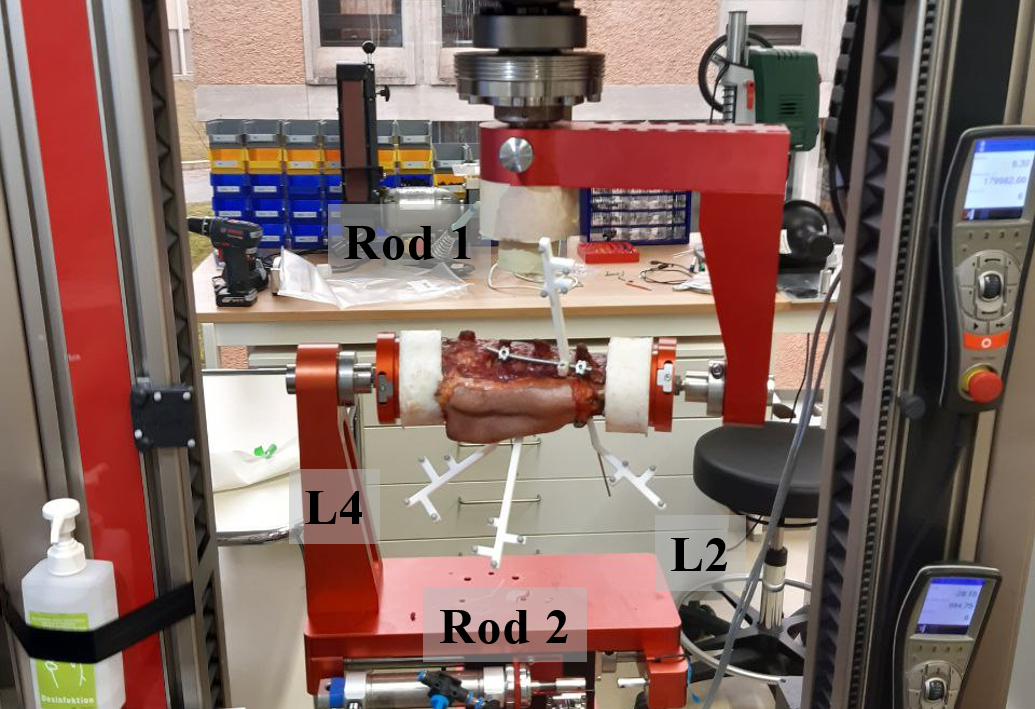}
    \caption{One of four cadaveric human lumbar spine specimens fixed in the biomedical testing machine. Individual passive infrared tracking markers are attached to vertebrae L2 and L4, as well as to the two rods of the implant.}
    \label{fig:zwick}
\end{figure}

To validate the simulated screw loosening, we mounted each specimen in a biomechanical testing machine ZwickRoell Z010 (ZwickRoell GmbH \& Co. KG, Ulm, Germany) which allows a defined and reproducible flexion-extension movement of the anatomy. To analyze the movement of the implants and the vertebrae, we attached passive optical tracking markers to the two rods, as well as to vertebrae L2 and L4. A high-fidelity optical tracking system, Atracsys fusionTrack 500 (Atracsys LLC, Puidoux, Switzerland), was used to record the trajectories of all tracked objects during movement. The experimental setup is shown in figure \ref{fig:zwick}.

The protocol of the biomechanical testing machine was programmed with a maximal torque of 7.5\,Nm in both directions, which determines the endpoint of the flexion-extension movement, and an angular speed of \SI{5}{\degree\per\second}. These values are standard settings for the biomechanical testing of the human lumbar spine according to empirical findings \cite{wilke1998testing}. For the experimental validation of screw loosening, we run 30 - 50 cycles of flexion-extension movement and record the tracking data of the rigidly attached markers.

For all implants and specimens, screw loosening was confirmed by an experienced spine surgeon who conducted all experiments. As an additional experimental quantification metric for screw loosening, we introduce a ratio defined as the relation between fixed and loose configurations of the relative movement between target vertebra and implant. In biomechanical screw toggling experiments, the level of screw loosening is usually measured as the relative displacement between implant and bone through optical tracking data as described in the work of Liebsch et al. \cite{liebsch2018pedicle}. However, these setups usually only include a single vertebra and screw mounted in a biomechanical testing machine which is only a simplified version of our setup. Furthermore, the proposed ratio compensates for subject-specific variations of relative displacement due to inter-subject bone quality differences.

First, we compute the relative movement $\Delta x$ of each respective rod and the vertebra L2 for all configurations, where $x$ is the reference point for each individual tracking target:

\begin{equation}
    \Delta x_i = || x_{L2} - x_{Rod_i} ||
\end{equation}

The centered mean absolute relative movement is computed as a scalar measure of the amount of relative movement between the implant and the vertebra L2, where $n$ is the number of synchronized measurements:

\begin{equation}
    \bar{x}_i = \frac{1}{n} \sum_{i=1}^{n} | \Delta x_i - \frac{1}{n} \sum_{i=1}^{n} \Delta x_i |
\end{equation}

Finally, we define the ratio of the relative movement between loose and fixed configuration as the loosening criterion and consider the screw as loose if the computed ratio exceeds a threshold of 2, which corresponds to a doubled relative movement of implant in regard to the target vertebra from fixed to loose configuration.

\begin{equation}
    R_{lf, i} = \frac{\bar{x}_{i, loose}}{\bar{x}_{i, fixed}} > 2
\end{equation}

\subsection{Vibro-Acoustic Sensing for Screw Loosening Detection}

\subsubsection{Experimental Setup}

\begin{figure}
    \centering
    \includegraphics[width=0.6\textwidth]{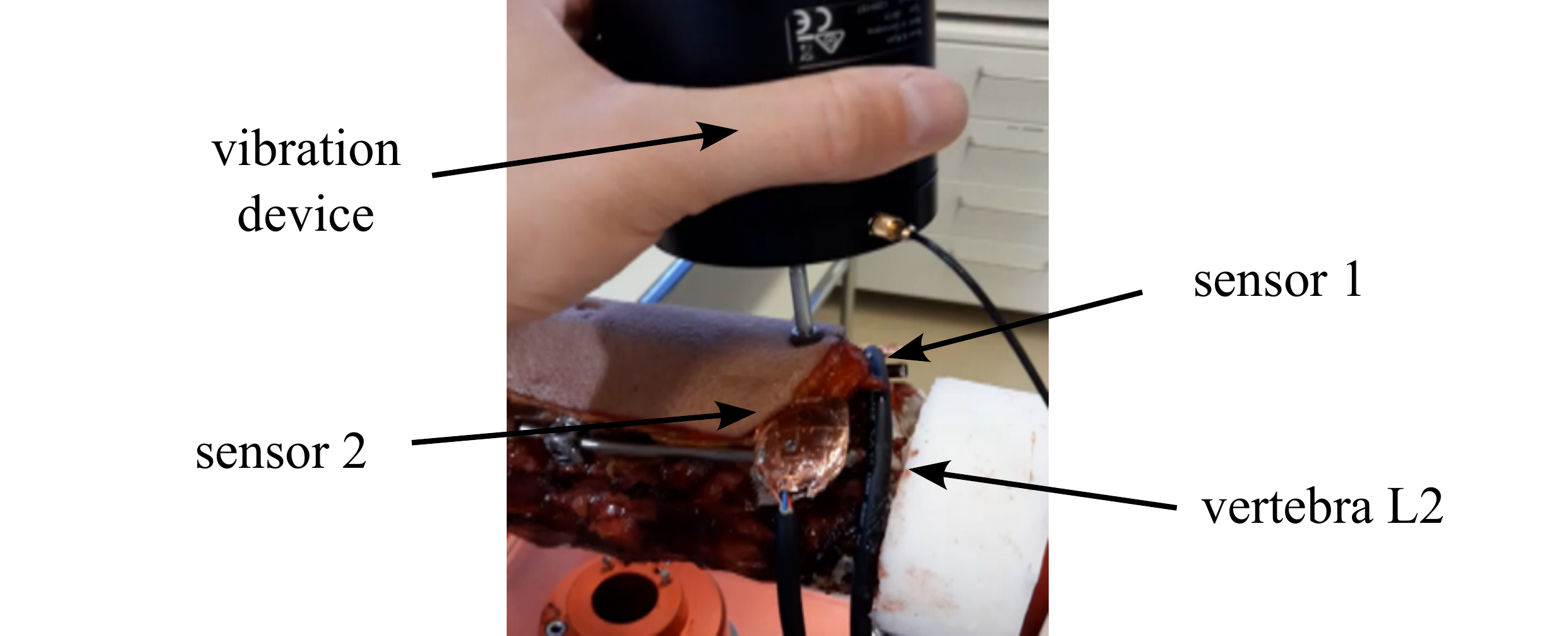}
    \caption{The vibration device is placed on top of the spinous process of the vertebra L2. The structure-borne vibrations are propagated through the bone into the screw shaft and measured with custom piezo-based contact microphones directly at the screw head.}
    \label{fig:vibration}
\end{figure}

For the detection of pedicle screw loosening, we apply active vibration excitation to the target anatomy. We place a vibration device (shaker type 4810, Brüel and Kjær, Teknikerbyen 28, DK-2830 Virum, Denmark) on the skin centered on top of the spinous process of the target vertebra L2 in an upright position and excite the tissue with a sine sweep in a frequency range of 10 to 500 Hz and with a duration of \SI{2.5}{\s}. To standardize the applied pressure and to make the system easily usable, we use the weight of the shaker (\SI{1080}{\gram}) to define the contact pressure on the skin of the specimen and hold it manually in place. We use a digital oscilloscope, Digilent Analog Discovery 2 (Digilent, 1300 NE Henley Ct. Suite 3, Pullman, WA 99163, USA), and its MATLAB (MathWorks, 1 Apple Hill Drive, Natick, MA, USA) interface to generate the excitation signal. We amplify (type 2706, Brüel and Kjær) the generated analog signal to drive the shaker. Figure \ref{fig:vibration} illustrates the experimental setup for the vibration experiments.

The vibration is propagated through the bony tissue of the vertebra to the screw shaft and recorded with custom piezo contact microphones which are directly glued to the screw head. The piezo contact microphones use a custom preamplification and impedance buffering stage to preserve low frequency content as described in \cite{seibold2021realtime}. We recorded a total number of 50 sine sweeps per screw and lifted and replaced the shaker device between the individual sweeps to have variation in the captured training data. We first measure every specimen in fixed configuration as illustrated in section \ref{sec:validation}, afterwards we intentionally loosen the screws of vertebra L2 using the protocol described in section \ref{sec:loosening} and repeat the whole procedure which results in 200 samples recorded per specimen. Using four human cadaveric lumbar spine specimens, we recorded a balanced dataset with a total number of 800 individual samples. All signals were captured in lossless wave file format, using a sample rate of \SI{44.1}{\kilo\Hz} and a bit depth of 24.

\subsubsection{Pedicle Screw Loosening Detection Algorithm}
\label{sec:detection}

\begin{figure}
    \centering
    \includegraphics[width=0.9\textwidth]{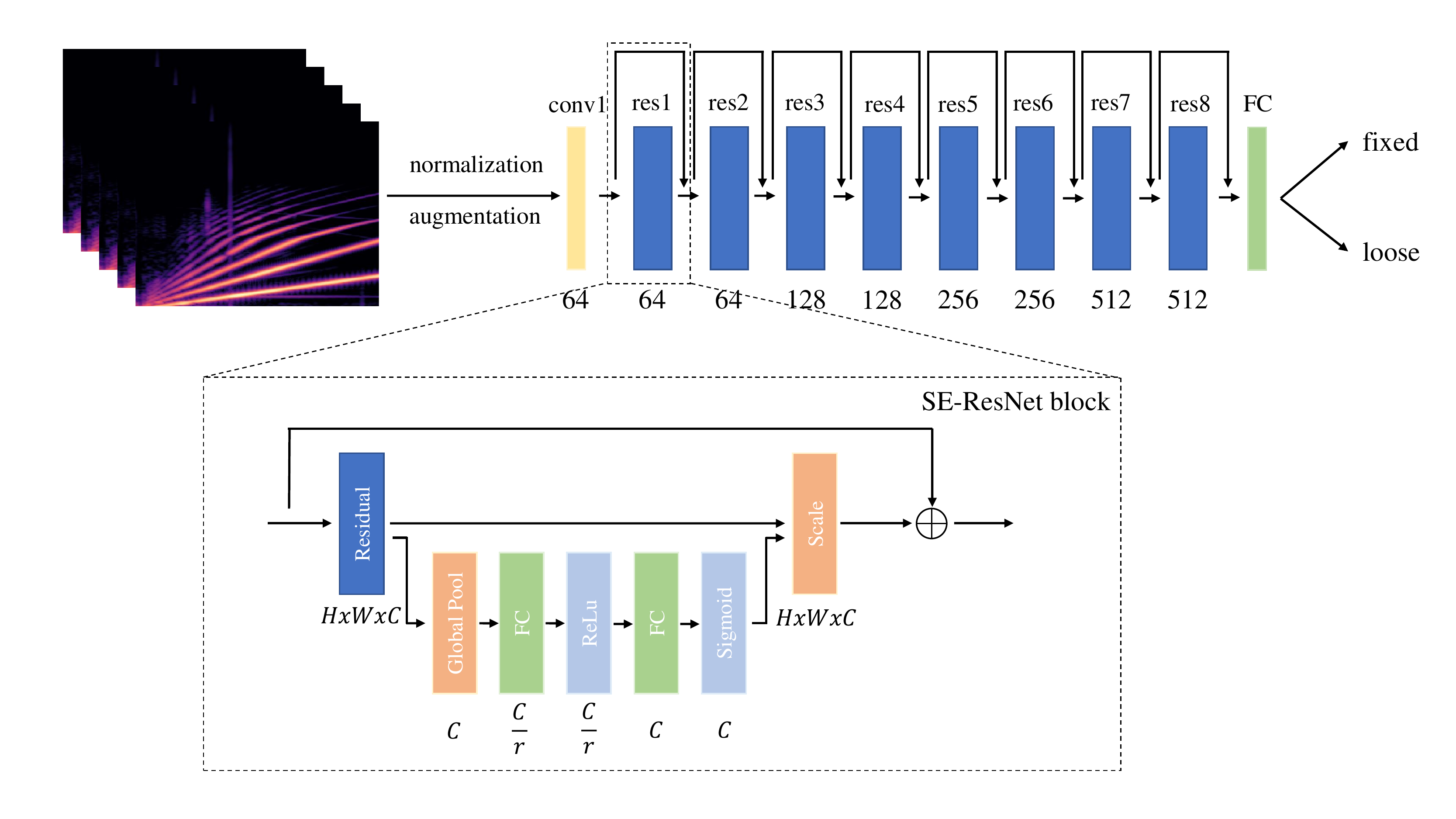}
    \caption{The overview of the proposed pedicle screw loosening detection pipeline. The spectrograms are fed to an 18 layer ResNet variant which implements Squeeze \& Excitation \cite{hu2018se} in each residual block. We define the detection as a binary classification problem, where the model is trained to differentiate between fixed and loose samples. In the SE-ResNet block schematic, the dimensions of each layer are illustrated where the variable $r$ stands for the reduction ratio as described in \cite{hu2018se}. The number of filters for each SE-ResNet block are given below, every layer employs a filter size of 3x3. The spectrograms are colorized for visualization purposes, however all spectrogram used in the implementation of this work are two-dimensional matrices with dimensions 256x218.}
    \label{fig:model}
\end{figure}

State-of-the-art systems in audio classification use a combination of log-mel spectrogram representation for the audio signal and a feature extraction backbone based on convolutional neuronal networks \cite{purwins2019deep,ciric2021spectrogram}. Therefore, we generate log-mel spectrograms with dimensions 256x218 from all individual samples in the dataset. The log-mel spectrograms serve as input for a modified 18-layer ResNet \cite{he2016resnet} and are computed using the python library librosa 0.7.2 \cite{mcfee2015librosa}. Log-mel spectrograms are two-dimensional matrices with time windows as columns, frequency mel-bins as rows, and amplitude as scalar matrix values. The first step to compute the log-mel spectrogram of an audio sample of length $N$ is to compute Short-Time Fourier Transformation (STFT):

\begin{equation}
	\label{eq:stft}
	X(m,k) = \sum_{n=0}^{N-1} x(n+mH)w(n) exp(\frac{-2\pi ikn}{N})
\end{equation}

We use the Hann window function as $w(n)$ to compensate for spectral leakage \cite{lyon2009spectral} and apply a hop length of $H = 256$. We map the resulting STFT $X$ which is structured as the $k^{th}$ Fourier coefficient (on the y-axis) for the $m^{th}$ time frame (on the x-axis) from amplitude to decibel by computing: 

\begin{equation}
	\label{eq:energy2db}
	X_{dB}(m,k) = 10 \, log_{10}(X(m,k)\textsuperscript{2})
\end{equation}

Finally, the spectrogram is mapped to the mel scale by applying a total number of 256 triangular filters which are evenly distributed on the Mel scale defined by:

\begin{equation}
	\label{eq:mel}
	f_{mel} = 2595 \, log_{10}(1 + \frac{f}{700})
\end{equation}

Compared to previous work, we extended the ResNet-18 backbone with Squeeze \& Excitation (SE) \cite{hu2018se} blocks which add a channel-wise attention mechanism to each residual block while introducing minimal computational overhead. This modification resulted in a substantial performance improvement in our experiments. In our implementation, we use a reduction ratio of $r = 8$ in all SE blocks. An overview of the proposed detection pipeline is illustrated in figure \ref{fig:model}.

The network is trained for 10 epochs using the Adam optimizer, a learning rate of $L_R = 1e-5$, and a binary crossentropy loss. We normalize all log-mel spectrograms $X_{norm, mel} = {(X_{mel} - \mu )} / {\sigma}$, where $\mu$ is overall mean and $\sigma$ is the standard deviation computed over the entire dataset. Furthermore, we augment the dataset by applying pitch shifting in the range of [-3, 3] semitones and time stretching with the factors [0.9, 1.1] to the raw waveforms directly. All experiments were implemented in Tensorflow/Keras 2.6 and executed on a NVidia RTX 2080 SUPER GPU.

\section{Results}

\subsection{Validation of Screw Loosening}

Table \ref{tab:displacement} shows the relative displacement of implant and vertebra L2 measured using the optical tracking system for every implant and specimen in both fixed and loose configurations. It can be observed that every specimen and implant shows an increased displacement after intentional loosening. To compensate for inter-subject variations in relative displacement due to differences in bone quality, anatomy and mounting in the biomechanical testing machine, we computed the loosening ratio for each individual screw as described in section \ref{sec:validation} as the main metric for the assessment of screw loosening. Table \ref{tab:validation} contains the computed ratios for all screws and specimens tested in our experiment. 

\begin{table}
    \centering
    \begin{tabular}{|c||c|c||c|c|}
        \hline
        \textbf{ID} & \textbf{Screw1 fixed} & \textbf{Screw1 loose} & \textbf{Screw2 fixed} & \textbf{Screw2 loose} \\ \hline\hline
        0 & 0.5998 mm & 2.5323 mm & 0.9927 mm & 3.3198 mm \\ \hline
        1 & 0.5122 mm & 1.3140 mm & 0.2181 mm & 1.5746 mm \\ \hline
        2 & 0.9234 mm & 2.0103 mm & 0.6141 mm & 6.2079 mm \\ \hline
        3 & 1.2149 mm & 3.3013 mm & 0.4763 mm & 4.8975 mm \\ \hline
    \end{tabular}
    \caption{The relative displacement of implant and target vertebra.}
    \label{tab:displacement}
\end{table}

\begin{table}
    \centering
    \begin{tabular}{|c|c|c|}
        \hline
        \textbf{ID} & $\mathbf{R_{lf, screw1}}$ & $\mathbf{R_{lf, screw2}}$  \\ \hline\hline
        0 & 6.9462 & 4.2048 \\ \hline
        1 & 4.7403 & 28.3358 \\ \hline
        2 & 2.6498 & 14.1524 \\ \hline
        3 & 3.1605 & 15.5463 \\ \hline
    \end{tabular}
    \caption{The values of $R_{lf,screw1}$ and $R_{lf,screw2}$ for all specimens and screws. The ratios have been computed according to the measurements and formulas described in sections \ref{sec:loosening} and \ref{sec:validation}.}
    \label{tab:validation}
\end{table}

\subsection{Screw Loosening Detection Results}

The spectrograms are analyzed using the detection pipeline proposed in section \ref{sec:detection}. To thoroughly evaluate the model performance, we perform a four-fold cross validation experiment and split the data on a specimen level. To this end, we train an individual model from scratch on the data collected from three specimens and test on the remaining specimen. All results are reported in the format \textit{mean} $\mathit{\pm}$ \textit{standard deviation}.

For the detection of pedicle screw loosening, our model reaches a sensitivity of $91.50 \pm 6.58 \%$ and a specificity of $91.10 \pm 2.27 \%$. These values correspond to a mean accuracy of the detection algorithm of $91.29 \pm 4.28 \%$. To give further insights, we report the results on the individual folds in table \ref{tab:kfold}, where the specimen ID corresponds to the specimen used for the test set, the model is trained on the remaining three specimens.

\begin{table}
    \centering
    \begin{tabular}{|c|c|c|}
        \hline
        \textbf{Specimen ID} & \textbf{Sensitivity} & \textbf{Specificity}  \\ \hline\hline
        0 & 94.00\% & 94.00\%  \\ \hline
        1 & 81.00\% & 87.76\% \\ \hline
        2 & 92.00\% & 90.62\% \\ \hline
        3 & 99.00\% & 92.00\% \\ \hline
    \end{tabular}
    \caption{Sensitivity and specificity reported for each individual fold in the four-fold cross validation experiment conducted for the evaluation of the proposed pedicle screw detection algorithm.}
    \label{tab:kfold}
\end{table}

We furthermore performed an ablation study to show the benefit of modifying the ResNet-18 backbone with Squeeze \& Excitation modules for the given problem. Without Squeeze \& Excitation modules, the model reached a sensitivity of $87.75 \pm 9.91\%$ and a specificity of $90.04\pm 7.19\%$ which corresponds to a mean accuracy of $88.89\pm 4.08\%$.

\section{Discussion}

In this work, we propose a detection method for pedicle screw loosening based on vibroacoustic sensing which could be an important step towards a novel radiation-free and non-invasive assessment method to improve the diagnostics in clinical practice and patient safety in revision surgery. We thoroughly evaluate our algorithm using k-fold cross validation and split the dataset on the specimen level. To the best knowledge of the authors, we propose the first alternative to medical imaging based assessment methods. Our experimental design may also allow clinical translation to a percutaneous application with reproduction of the typical pain in the context of symptomatic screw loosening. This would help physicians as the clinical correlation of radiological findings of screw loosening with the complained symptoms is not always evident. To address the aforementioned problem, the proposed learning-based pedicle screw loosening detection algorithm shows promising performance indicating great potential for the development of systems for the automated screw loosening detection based on vibroacoustics. As the target vertebra is excited with sine sweep vibration, the resulting measurements at the screw head are greatly influenced by the anchorage of the screw in the surrounding bone tissue. A fan-shaped cavity around the screw shaft therefore changes the transmitted vibration characteristics which serves as the basic structure-borne sound propagation mechanism that motivates our work. 

3D-printed surgical guides were introduced as an approach for a more time efficient simulation of pedicle screw loosening which was confirmed to be sufficiently realistic through the analysis of optical tracking data. Extended simulation with toggling experiments would probably result in a more realistic screw loosening model as the loosening funnel was uniformly designed for all specimens. However, the focus of the present work is the development of a vibroacoustic-based method for screw loosening detection, the implementation of a highly realistic loosening simulation is not in the scope of this work and should be investigated in future research. Furthermore, as we fully loosened the screw in our experiments, the influence on the detection performance of the proposed algorithm with different levels of screw loosening has to be investigated in future work. In addition, the influences of patient body mass index (BMI) and bone quality on the proposed system should be taken into consideration.

After conducting the loosening simulation process as described in section \ref{sec:loosening}, an experienced spine surgeon confirmed the loosening of the respective pedicle screws visually and haptically. To additionally quantify the screw loosening, we introduced a loosening ratio which is computed using optical tracking data. Hereby, the relative displacement of implant and target vertebra shown in \ref{tab:displacement}, as well as the loosening ratio shown in table \ref{tab:validation} show certain variations. The reasons for these variations are subject-specific bone quality and  anatomical differences. Furthermore, it is practically unfeasible to install the specimen perfectly centered in the biomechanical testing machine which results in a slightly asymmetric movement. However, the loosening validation experiments show a more than doubled relative movement between implant and respective vertebra which can be considered sufficient for simulating pedicle screw loosening in our experiments.

A limitation of the presented study is the small sample size of four human cadaveric specimens. However, by performing a four-fold cross validation experiment and showing the consistency of the results over all four individual test folds, we consider our experiment as a strong proof-of-concept for the usage of vibroacoustic sensing in orthopedics. The variations in the per-fold model performance can be accounted to anatomical variations and bone qualities. Additional \textit{in-vitro} and \textit{in-vivo} studies have to be performed to test the reliability of the system and increase the training data for better generalization and detection performance. 

With a shaker device that applies the vibration on the patient's skin over the spinous process to the bone and using its weight as contact force, we propose an easy-to-integrate measurement method which requires only little additional human effort. In the present work, we chose a bilateral posterior approach for the surgical access, however, also with a central surgical access, the method is suitable for intraoperative detection of loose pedicle screws, as the vibration device cannot only be placed on top of the spinous process on the skin, but also directly on the bone. In future, we envision the presented approach not only to be valuable as an intraoperative confirmation of pedicle screw loosening, but also as a clinical tool for the preoperative diagnosis of screw loosening and, eventually, as the foundation to design smart pedicle screws to monitor or even predict pedicle screw loosening in a reliable and non-invasive way. Nevertheless, additional research and development is required to design custom sensorized implants, transmit the signals to the outside of the human body and solve the energy supply.

\section{Conclusions}

We propose a non-destructive, radiation-free and easy-to-integrate approach to detect pedicle screw loosening intraoperatively using active vibroacoustic sensing. The resulting system could be used for the intraoperative confirmation of loose pedicle screws as an alternative for the measurement of the extractional torque. Furthermore, we believe that the proposed work could be a strong proof-of-concept for the development of smart implants for spinal fusion surgery.

\section*{Acknowledgments}

This work is part of the SURGENT project and was funded by University Medicine Zurich/Hochschulmedizin Zürich.


\bibliographystyle{elsarticle-num} 
\bibliography{bibliography}

\end{document}